# Charge photogeneration in few-layer MoS$_2$

*Tetiana Borzda[1,2,]\*, Christoph Gadermaier[1,2,]\*, Natasa Vujicic[1,3], Peter Topolovsek[1,2], Milos Borovsak[1], Tomaz Mertelj[1], Daniele Viola[4], Cristian Manzoni[4], Eva A. A. Pogna[4], Daniele Brida[5], Maria Rosa Antognazza[6], Francesco Scotognella[4,6], Guglielmo Lanzani[4,6], Giulio Cerullo[4], and Dragan Mihailovic[1,2,7]*

[1]Department of Complex Matter, Jozef Stefan Institute, Jamova 39, 1000 Ljubljana, Slovenia

[2]Jozef Stefan International Postgraduate School, Jamova 39, 1000 Ljubljana, Slovenia

[3]Institute of Physics, Bijenicka 46, 10000 Zagreb, Croatia

[4]IFN-CNR, Department of Physics, Politecnico di Milano, P. Leonardo da Vinci 32, 20133 Milan, Italy

[5]Department of Physics, University of Konstanz, PO box M696 78457 Konstanz, Germany

[6]Center for Nano Science and Technology, Italian Institute of Technology, Via Pascoli 70/3, 20133 Milano, Italy

[7]Center of Excellence in Nanoscience and Nanotechnology, Jamova 39, 1000 Ljubljana, Slovenia

ABSTRACT. The two-dimensional semiconductor MoS$_2$ in its mono- and few-layer form is expected to have a significant exciton binding energy of several 100 meV, leading to the consensus that excitons are the primary photoexcited species. Nevertheless, even single layers show a strong photovoltaic effect and work as the active material in high sensitivity photodetectors, thus indicating efficient charge carrier photogeneration (CPG). Here we use continuous wave photomodulation spectroscopy to identify the optical signature of long-lived charge carriers and femtosecond pump-probe spectroscopy to follow the CPG dynamics. We find that intitial photoexcitation yields a branching between excitons and charge carriers, followed by excitation energy dependent hot exciton dissociation as an additional CPG mechanism. Based on these findings, we make simple suggestions for the design of more efficient $MoS_2$ photovoltaic and photodetector devices.

Recent progress in the exfoliation of layered materials[1,2] and the nanofabrication of functional structures has revived the interest in two-dimensional materials with properties complementary to graphene, in particular transition metal dichalcogenides[3,4] (TMDs) such as $MoS_2$. Depending on the metal atoms' coordination and oxidation state, TMDs can be metallic, semimetallic, or semiconducting. Additionally, some TMDs show superconductivity[5], charge-density waves[6] and hidden electronically ordered phases[7]. Their potential for electronics has become evident by the realization of a field effect transistor[8] (FET) and a logic circuit device[9] based on a single flake of monolayer $MoS_2$.

The optical absorption of $MoS_2$ in the visible spectral range shows four excitonic resonances[10], commonly labeled (see Figure 1a) A to D at 1.9, 2.1, 2.7, and 2.9 eV. The spectral positions of these resonances are almost independent of the number of layers, while the indirect band gap is at 1.2 eV in the bulk and grows progressively as the number of layers is reduced, even exceeding the energy of the A exciton resonance for the monolayer. Hence the monolayer, contrary to bi- and multilayers, behaves like a direct gap semiconductor and shows significant fluorescence[11,12]. The exciton binding energy for bulk $MoS_2$ has been determined to be 45 meV and 130 meV for the A and B excitons, respectively[13]. Both exciton binding energies increase upon decreasing the sample thickness, with estimates for monolayers[14-16] ranging from 0.4 to 0.9 eV. Despite this high exciton binding energy, monolayer $MoS_2$ shows a strong photovoltaic effect[17] and potential for high sensitivity photodetectors[18]. Both findings require efficient charge carrier photogeneration (CPG), either via direct excitation of mobile carriers or via exciton dissociation.

The spectral signature of charge carriers has been identified by absorption and fluorescence spectroscopy of $MoS_2$, where the charge concentration is varied either via the gate voltage in a FET geometry[19] or via adsorption[20] or substrate doping[21]. The absorption peaks of charges are red-shifted by about 40 meV compared to the ground state absorption into the A and B excitons and have been attributed to optical transitions from a charged ground state to a charged exciton (trion). The possibility of alternative interpretations, such as polarons[22,23] or Stark effect in the local electric field of the charges[24-26] does not compromise the identification of these absorption peaks as belonging to charges.

Here we use continuous wave (cw) photomodulation (PM) and femtosecond pump-probe spectroscopy to identify the spectral features of photogenerated charges and trace their dynamics, starting with their generation either by direct impulsive excitation into the charge continuum or via exciton dissociation. We exfoliated $MoS_2$ in ethanol, following the protocol in Ref 1. The dispersion was dried and the obtained few-layer flakes were re-dispersed in a solution of PMMA, a transparent and electronically inert polymer. This dispersion was spin-cast onto a quartz substrate, yielding a macroscopic PMMA film with a homogenous greenish-yellow color characteristic of thin $MoS_2$ films[27] (see Fig. 1a). PMMA in this sample serves as a matrix that holds an ensemble of flakes with lateral size1 of few-hundred nm. More details about the sample preparation are found in the Methods section. The distance $\Delta\omega$ between the two Raman peaks around 400 cm$^{-1}$ is generally viewed as the most robust measure of the flake thickness[28]. From the Raman spectra in Fig 1b we obtain $\Delta\omega = 23$ cm$^{-1}$ for excitation at 633 nm and $\Delta\omega=25$ cm$^{-1}$ at 488 nm. Within the distribution, various thicknesses contribute differently at the two excitation wavelengths[29]. Overall, the Raman spectra indicate a flake thickness distribution that is dominated by three- to six-layer flakes.

The advantages of this kind of sample compared to individual flakes are the ease of fabrication and handling and the possibility to use any spectroscopic technique without the need for high-resolution optical microscopy. $MoS_2$ flakes embedded in PMMA are in a slightly different environment than mono- or few layer flakes on dielectric substrates used in previous femtosecond studies[30,31]. However, as our results will show, the spectra and the relaxation times of the signal are very similar to those obtained on individual few-layer flakes. Hence the present study directly extends existing knowledge on the femtosecond behavior of few-layer $MoS_2$.

Results and Discussion

The absorption spectrum in Fig. 1a shows the characteristic A and B exciton resonances, which are broader and red-shifted compared to undoped $MoS_2$, as is typical of commercial $MoS_2$ of mineral origin[32], which is doped due to dislocations induced by the exfoliation and due to (mostly metallic) impurities. Hence, each of the two absorption peaks is actually an overlap of at least two contributions: neutral ground state to exciton absorption at the higher energy side of each peak and lowest charged state to excited charged state at the lower energy side. For the A peak, an even lower energy contribution has been identified[32], which has alternatively been ascribed to a surface trapped exciton[33], an edge state[34], or a plasmon resonance[35], so that peak A actually arises from the overlap of three peaks. Similar to the notation in Refs 19, 20, and 32, we will use the labels L, $A^-$, and $A^0$ for the low energy peak, charge peak and exciton peak of the A resonance, and $B^-$ and $B^0$ for the charged and neutral contributions to the B resonance. Please note that, contrary to electrical or chemical doping, photoexcitation generates charges in pairs of opposite polarities. However, although we expect the signatures of the corresponding positive charges at the same spectral positions, we know only those of the negative charges, hence our labeling.

To identify long lived photoexcitations we performed cw PM spectroscopy. Here, a cw laser with 3.1 eV photon energy used for exciting the sample is periodically modulated via a mechanical chopper. The relative change $\Delta T/T$ of the transmitted light from a halogen lamp is measured via phase sensitive detection (see Methods section for details). Those photoexcitations whose population changes significantly over the modulation cycle (i.e. their lifetime is long enough to build up sufficient population while the laser is on and short enough to sufficiently reduce their population while it is off) are identified in the cw PM spectrum via their photoinduced absorption (negative $\Delta T/T$) transitions to higher excited states. Concomitantly with the increase of photoexcited populations, the ground state population and its associated absorption is reduced (photobleaching, positive $\Delta T/T$). The cw PM spectrum at room temperature upon excitation at 3.1 eV, above the C and D exciton resonances, is shown in Fig. 2a. The signal is largely in phase with the modulation of the photoexcitation, with a negligible quadrature contribution. This means that the populations at the origin of the signal can easily follow the modulation at 245 Hz, implying that the lifetimes of the respective photogenerated species are much shorter than the modulation period of ~4 ms. The main features of the spectrum are three positive (photobleaching) and two negative (photoinduced absorption) peaks. We fit the spectrum using five overlapping Gaussians (see Fig 2b), which represent the thermal and disorder-induced (in particular by polydispersity of flake thickness) broadening of the electronic resonances. The strong overlap between neighboring peaks makes them appear narrower than their actual lineshape and connected by an almost straight line, masking the inflection points characteristic of isolated Gaussian peaks. The spectral positions of the five peaks correspond very well with the three neutral and two charge peaks discussed in References 19, 20, and 32. Assuming the same origin for the peaks in the PM spectrum, we obtain a straightforward interpretation. Upon photoexcitation, the number of electrons in the neutral ground state is reduced, and the number of charge carriers is increased. Hence the absorption features L, $A^0$, and $B^0$ from transitions between neutral states are reduced, resulting in a positive $\Delta T/T$ (photobleaching), while the absorption features $A^-$ and $B^-$ from charges are increased, yielding a negative $\Delta T/T$ (photoinduced absorption).

Alternatively, one could interpret the PM spectrum based on its resemblance of a derivative lineshape. A photoinduced blue shift of the absorption spectrum would result in a $\Delta T/T$ contribution that follows the first derivative of the absorption spectrum (or a negative first derivative for a red shift); a photoinduced broadening of the absorption peaks would contribute a negative second derivative. If we interpreted our spectrum in terms of derivative lineshapes, it would be dominated by a positive second derivative, which indicates a photoinduced line narrowing. We are not aware of any such mechanism. However, the A peak is composed of the three overlapping narrower peaks L, $A^-$, and $A^0$ (no equivalent to the L peak has yet been identified for the B peak, but we may extrapolate our reasoning also to B). In our proposed scenario photoexcitation generates charges and the middle peak A$^-$ increases at the expense of the other two, which decreases the overall width of the A peak. Hence, CPG leads to an apparent photoinduced line narrowing, which explains the positive second derivative lineshape.

To investigate CPG in real time, we now turn to femtosecond optical pump-probe spectroscopy. Like cw PM, this technique measures the relative change in transmission $\Delta T/T$. However, rather than continuously, the sample is photoexcited at a well-defined point in time by a fs laser pulse (the pump) and the transmission spectrum is measured with a second fs laser pulse (the probe) at a well-defined delay after the pump. Scanning the pump-probe delay allows to follow the evolution of the photoexcited states' populations (see details in the Methods section). We start by comparing the $\Delta T/T$ signal for excitation at 3.1 eV at long pump-probe delay (300 ps ) with the cw PM (see Fig. 3a). The two normalized spectra are very similar, with three important differences: the L peak is absent, there is an additional broad photoinduced absorption feature peaking around 2.45 eV, and the whole spectrum is red shifted. The red shift of the whole spectrum is more pronounced at higher pump intensities (see Fig 3b) and shorter pump-probe delays (see Figs 3c and d). Both these correlations suggest that the red shift is stronger for higher concentration of a certain species of excited states. The most intuitive interpretations are Stark effect due to the local field of photogenerated charges, as has been observed in semiconductor nanocrystals[24], organic semiconductors[25], and carbon nanotubes[26,36], inter-excitonic interaction[37] or band gap renormalization, as has been found in semiconducting quantum wells upon photoexcitation[38] and inferred in recent works on semiconducting TMDs[39,40]. This intensity dependent red-shift also explains how the non-linear optical properties of MoS$_2$ can change from saturable absorption (i.e. photobleaching) to optical limiting (i.e.

photoinduced absorption) as a function of pump intensity[41]. After approximately 3 ps, the spectrum decays without any significant shifts or changes of shape, through a dominant process with a time constant of approximately 500 ps (see Fig. 3e), as previously obtained on few-layer $MoS_2$ supported on a dielectric substrate[30].

In addition to the previously identified $A^-$, $A^0$, $B^-$, and $B^0$ peaks, we note an additional broad absorption peak and a further positive peak at higher probe energies. Due to its position, we straightforwardly assign the positive peak to bleaching of the C exciton and label it $C^0$. The absorption peak shows a formation similar to $A^-$ and $B^-$ (see next paragraph), and is similarly long lived, hence it should belong to a charge population. Like the C exciton bleaching, it is absent in cw PM, and strongly reduced for excitation energies below the C exciton resonance, see (Fig 3f), hence we ascribe it to a charge associated with the C exciton, labeled $C^-$. The C exciton is ascribed either to excitation from a band below the valence band, or between the valence and conduction bands in a region of the bands nesting slightly off the $\Gamma$ point[42]. In both scenarios it is plausible that the $C^0$ and $C^-$ feature do not appear in the cw experiment, which probes populations that have relaxed towards the edges of the band gap.

During the first 3 ps the $\Delta T/T$ spectrum undergoes a characteristic change of shape: the positive signal components are formed during the instrumental resolution and decay monotonically, while the main photoinduced absorption features, associated to charges $A^-$, $B^-$ and $C^-$, show an initial instrument-limited rise followed by a delayed rise component after the pump pulse and by a slower decay. We can interpret this dynamics by assuming that the pump pulse creates an ensemble of excitons and charges with a combined spectrum $S_1(E)$. During the first 3 ps, this ensemble evolves into one with less excitons and more charges, with a combined spectrum $S_2(E)$. Subsequently, the exciton and charge populations decay with very little further change of shape, meaning that either the decay times of excitons and charges are very similar (either by coincidence or due an interdependent relaxation mechanism), or that the population after 3 ps is already dominated by charges. The latter hypothesis is corroborated by the similarity between the fs $\Delta T/T$ spectrum at longer delays and the cw PM spectrum, which should not show any exciton contribution. In the simplest possible case, the ensemble with the spectrum $S_1(E)$ evolves into the one with the spectrum $S_2(E)$ with only one characteristic sub-picosecond relaxation time $\tau_1$. Hence we propose the simple scheme of the underlying photoexcitation dynamics shown in Fig. 4a: The pump pulse generates a mixed population of charge carriers and excitons. Subsequently a certain fraction of the excitons dissociates into charges.

The evolution of the photoexcited states' population according to Fig. 4a is best fitted with a characteristic time $\tau_1 = 680$ fs. The assumption of only one common time constant $\tau_1$ for the dissociation of the A, B, and C excitons may be a gross simplification, but it describes the data remarkably well (Figs 4b+c). The spectra $S_1$ and $S_2$ in Figs. 4d and e can be fitted with overlapping Gaussians analogously to the cw PM spectrum. Besides the six already identified charge absorption and exciton bleaching features, there are two additional peaks, which we label X and Y. Since they do not show the delayed formation characteristic of charges, we assign them to photoinduced absorption by one of the exciton populations.

Comparing the relative contributions of the X and Y peaks to the $S_1$ and $S_2$ spectra, we can deduce that after 3 ps approximately the exciton population is somewhat decreased (by approximately a factor of 2). On the other hand, if the exciton population followed a curve $\sim exp(-t/\tau_1)$, the remaining exciton population would be only 1%. Hence, $\tau_1$ is not the time constant of the exciton dissociation, but rather the characteristic time with which the dissociation rate diminishes. Besides its time dependent rate, the exciton dissociation yield depends also on the energy of the exciting photons. This is expressed in Fig. 4a as a rate constant $k_d(E, t)$ that depends on the pump energy and on time. Fig 3f shows that the ratio between the absorption peaks due to charges and the respective bleaching peaks is higher for higher photon energy. This is consistent with the lower PL quantum yield for higher excitation energy, which has also been ascribed to charge separation[42].

Detailed studies of the CPG dynamics exist for carbon nanotubes and conjugated polymers, which are materials with exciton binding energies similar to $MoS_2$. In carbon nanotubes, there is an initial branching between mostly excitons and 1-2% directly excited charges[36], with a higher charge yield for higher excitation energy. In conjugated polymers, there is a similar initial branching, followed by additional CPG via dissociation of "hot" excitons during the first few picoseconds[43]. Elaborate models describe how the surplus energy of hot excitons increases their dissociation probability[44,45]. The relaxation of the electron and hole to the lowest exciton state are typically faster than our observed 680 fs[46-48]; however electron-phonon coupling creates a phonon heat bath that can live on for a few ps[49]. Additionally, exciton migration, which is facilitated by the extra energy and comes to a halt when the exciton reaches a local energy minimum, increases the probability that the exciton reaches a site where its dissociation is facilitated. In $MoS_2$ flakes, such sites could be surface defects, flake edges, metallic inclusions, crystal faults or small islands of an extra $MoS_2$ layer. We therefore conclude that the exciton dissociation probability is high while the excitons are hot and mobile and decreases as they reach their energetic minima. CPG in few-layer $MoS_2$ is a combination of two processes, both which have increased efficiency for higher exciting photon energy: direct excitation of charge pairs (within the 50-100 fs instrument resolution of our experiment) and hot exciton dissociation.

Before we discuss the implications of time- and pump energy dependent hot exciton dissociation, we review how our findings compare to previous femtosecond work on TMDs. Intervalley scattering[50,51] requires circular polarization and is not probed in our experiment, because all our laser polarizations are linear. On a time scale of 1 ps to 500 ps, Shi et al. obtained results very similar to ours in few layer flakes deposited on a dielectric substrate[30], but they do not discuss the temporal change of shape of the spectrum, which is crucial in understanding exciton dissociation and charge generation. They estimate that for excitation fluences similar to ours, the sample temperature should change by only 0.1 K, thus dismissing sample heating as a possible source of the signal. In our samples, due to the low heat conductivity of PMMA[52], heating could be a bit stronger. However, a signal due to heating would not change its shape with time and, most crucially, its shape would not depend on the excitation photon energy. In Fig 3f, there is a small $C^-$ signal even for excitation below the C exciton resonance. This contribution, which at any delay is 20% or less compared to the spectrum for excitation above the C resonance, may originate from heating or any other mechanism that changes the overall lineshapes and/or peak positions, such as Stark effect or band gap renormalization. Hence, for the combination of all such processes including heating, we estimate an upper boundary of 20% contribution. Any such mechanism is less important in the cw PM experiment, where no appreciable $C^-$ signal is found. Therefore, both the femtosecond and the cw signals are dominated by changes in excited and ground state populations, not pure lineshape/shift mechanisms.

Considering further lineshape/shift mechanisms, inter-exciton interaction is expected to result in a red-shift[37] and line broadening. However, we observe an apparent line narrowing (see discussion of the cw PM spectrum) due to the increase of the central $A^-$ peak at the expense of its neighbors on either side. The Burrstein-Moss effect, which has been observed in substitutionally doped $MoS_2$ fullerenes[53] should lead to a blue shift with increasing excited state population. Our contrasting observation implies that any Burrstein-Moss contribution is overwhelmed by one or more red-shifting mechanisms.

Conclusion and outlook

We have shown that charge carrier photogeneration in few layer $MoS_2$ arises from two different processes. First, there is a branching into excitons and charge carriers as the primary photoexcited species. Additionally, there is an increased, excitation energy dependent charge carrier yield from hot exciton dissociation during the first few ps. For monolayer $MoS_2$, due to the higher exciton binding energy[14], we can expect both CPG processes to have a lower yield. According to our findings, the efficiency of $MoS_2$ photovoltaic[17] and photodetector[18] devices depends significantly on the excitation wavelength and can be strongly increased, especially in monolayer devices, by facilitating exciton dissociation, e. g. via a strong built-in field using appropriate electrode materials, by engineering a p-n junction[54,55], or by combining $MoS_2$ with a second material[56-58] so one of them acts as electron donor and the other as acceptor in a heterojunction.

Methods:

MoS$_2$ sample preparation: 1 mg of MoS$_2$ in water/ethanol mixture (Graphene Supermarket) was extracted by flocculation with the addition of KCl (Potassium Chloride ≥99.0 %; Sigma) into a solution. Obtained flocculates were repeatedly washed with a fresh water/ethanol (1:1) mixture to remove any salt residues. Solvent mixture of water and ethanol was then replaced by the absolute ethanol only. Sedimented MoS$_2$ was extracted from the bottom with a minimum amount of solvent and redispersed in 2ml of chlorobenzene in ultrasonic bath. 40 mg of PMMA (Poly(methyl methacrylate) avg. M$_W$~350,000; Aldrich) was added into a solution of MoS$_2$ and sonicated at 50 °C for 30 min. 30 μl of the obtained stable solution was drop-casted onto a quartz substrate and left to dry in air.

All spectroscopic investigations were performed at room temperature. Raman spectroscopic characterization of MoS$_2$ was performed with an NT-MDT NTEGRA SPECTRA confocal Raman microscope in backscattering geometry with spectral resolution of 0.7 cm$^{-1}$. We used the excitation lines at 488 nm (Argon ion laser) and 632.8 nm (He-Ne laser) and a 100× objective (NA 0.9) to focus onto a spot size of 3microns. The Raman signals were detected with a CCD array at -70 °C. We used a laser power below 3 mW to avoid damage of the sample.

The probe used in the cw PM experiment is a halogen lamp (ASB-W-30 from Spectral Products) filtered by a monochromator (CM 112 from Spectral Products with 0.3 mm slits). Excitation is provided by diode laser at 405 nm (3.1 eV) with a power of 85 mW, focused onto a spot of 8 mm diameter, mechanically modulated by a chopper (MC 1000A Optical Chopper System from Thorlabs).

Ultrafast spectroscopy: the femtosecond pump-probe spectroscopy setup is driven by an amplified Ti:sapphire laser (Coherent Libra) producing 4-mJ, 100-fs, 1.55-eV pulses at 1-kHz repetition rate. A fraction of the pulse energy is used to drive a second-harmonic (SH) pumped OPA, generating ≈70-fs pulses tunable from 1.65 to 2.5 eV; the pump pulses are provided either by the OPA or by the SH of Ti:sapphire. Another fraction of the pulse energy is focused in a 3-mm-thick sapphire plate to generate a single-filament white light continuum used as a probe. Pump and probe are non-collinearly focused on the sample and the transmitted probe spectrum is detected by a spectrometer working at the full 1-kHz repetition rate of the laser. $\Delta T/T$ spectra are recorded with a time resolution of ≈100-fs and a sensitivity of $1 \div 2 \times 10^{-5}$.

Acknowledgements

We appreciate the stimulating discussions with V. Kabanov, I. Madan, D. Polli, J. Rehault, G. Seifert, A. Shumilin, and R. Tenne. The research leading to these results has received funding from LASERLAB-EUROPE (grant agreement no. 284464, EC's Seventh Framework Programme), Zukunftskolleg, the Marie Curie CIG project "UltraQuEsT" n° 334463, the ERC Advanced Grant "Trajectory" and the Graphene Flagship (contract no. CNECT-ICT-604391). The research was carried out in the context of the Marie-Curie ITN "MoWSeS" and the COST "Nanospectroscopy".

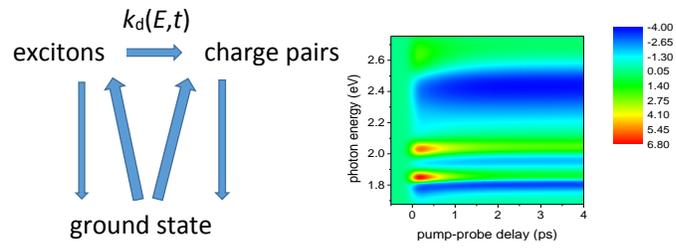

**Graphical Table of Contents.**

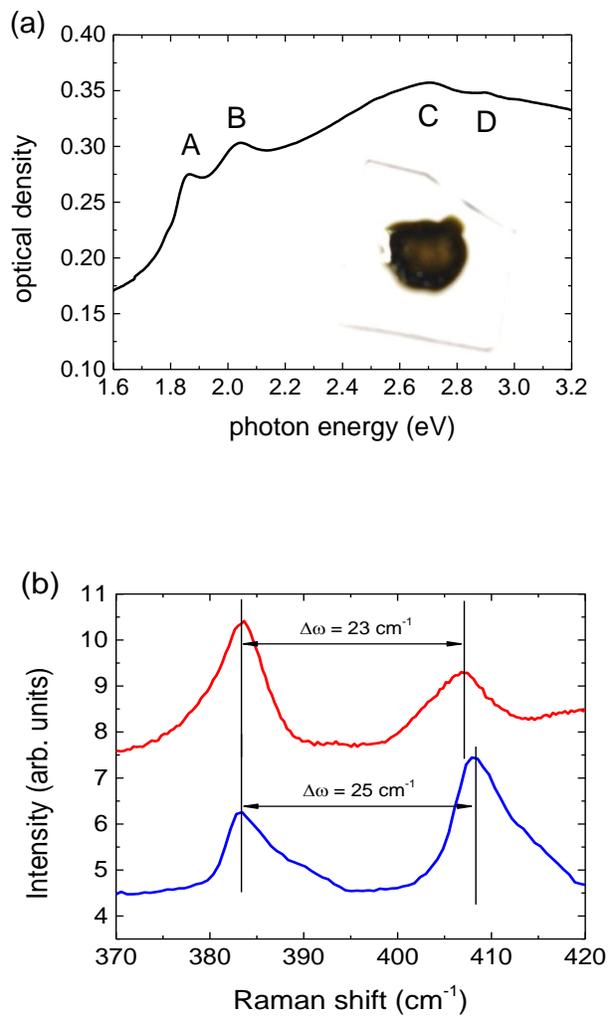

**Figure 1.** (a) Absorption spectrum of the sample of MoS$_2$ in PMMA. Inset shows a photograph (the dark area has a diameter of approximately 7-8 mm). (b) Raman spectra for two excitation wavelengths at 633 nm (red line) and 488 nm (blue line).

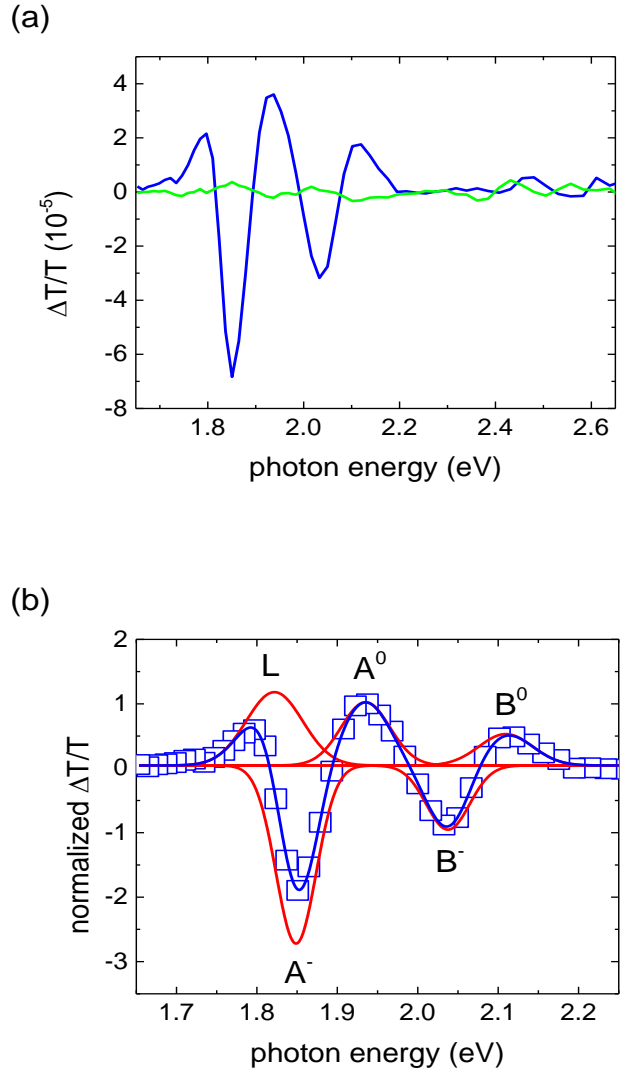

**Figure 2.** (a) cw photoinduced absorption spectrum of $MoS_2$ in PMMA at room temperature for excitation at 3.1 eV. In-phase (blue) and quadrature (green) signal components are shown for a modulation frequency of 245 Hz. (b) Five Gaussian fits (red) whose sum (blue) fits the normalized in-phase spectrum (open squares).

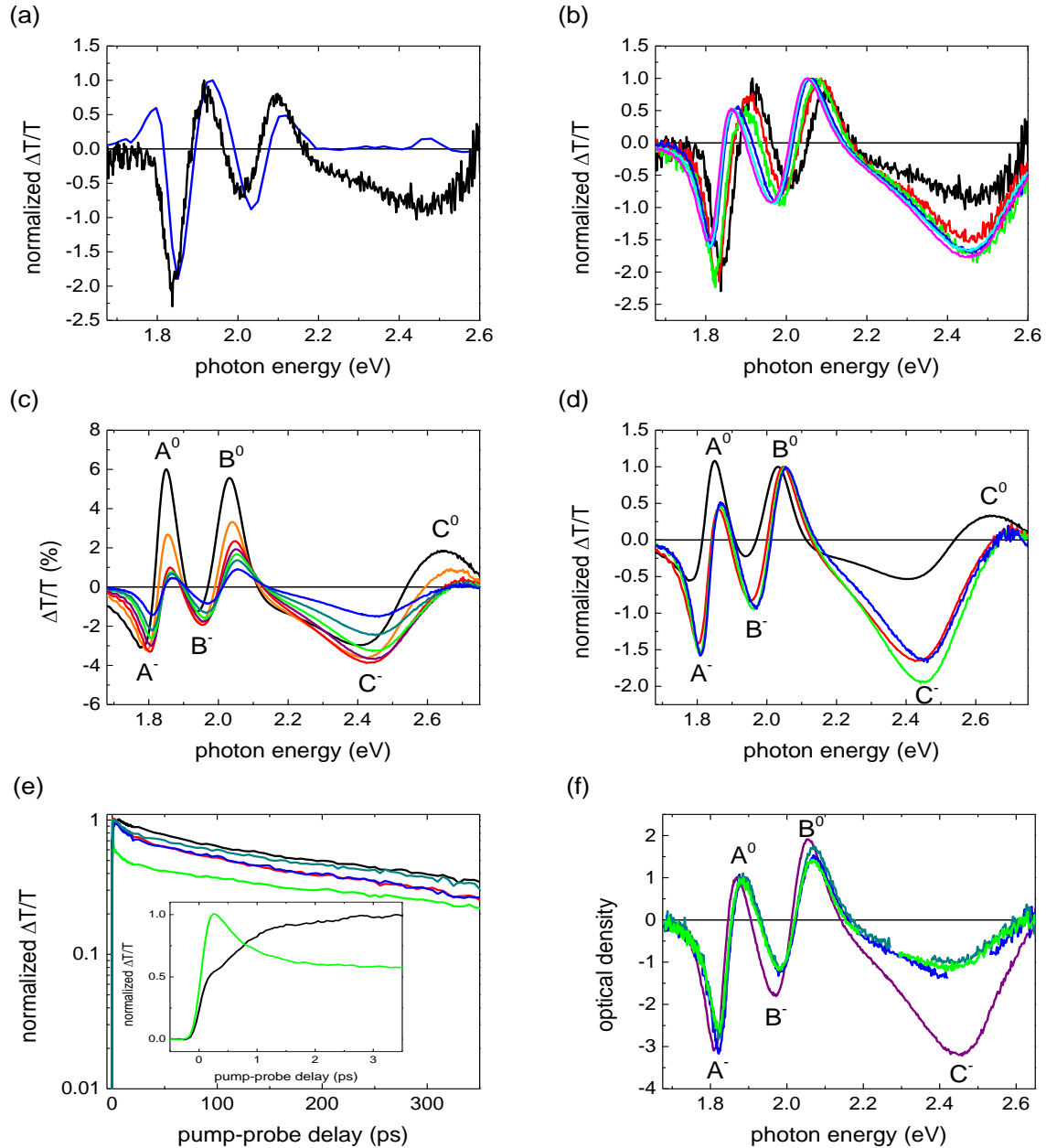

**Figure 3.** (a) Normalized cw photomodulation spectrum of MoS$_2$ in PMMA (blue) compared to the fs transient spectrum at 300 ps pump-probe (black) delay for 2 10$^{14}$ cm$^{-2}$ excitation fluence, 3.1 eV pump photon energy (b) pump-probe spectra at 300 ps normalized to the B exciton peak for different pump fluences at 3.1 eV pump photon energy: 2 (black), 3(red), 6 (green), 20 (blue), 40 (cyan) and 80 10$^{14}$ cm$^{-2}$ (c) absolute and (d) normalized (to the B exciton peak) pump-probe spectra for 4 10$^{15}$ cm$^{-2}$ pump fluence at delays 300 fs (black), 1 ps (orange), 3 ps (red), 10 ps (purple), 30 ps (green), 100 ps (dark cyan), and 300 ps (blue) (e) normalized time traces for different probe energies: 2.48 (black), 2.25 (red), 2.07 (green), 1.94 (blue) and 1.80 eV (dark cyan) (f) normalized spectra at 300 ps pump-probe delay for different pump photon energies: 3.10 (purple), 2.48 (blue), 2.34 (dark cyan) and 2.25 eV (green).

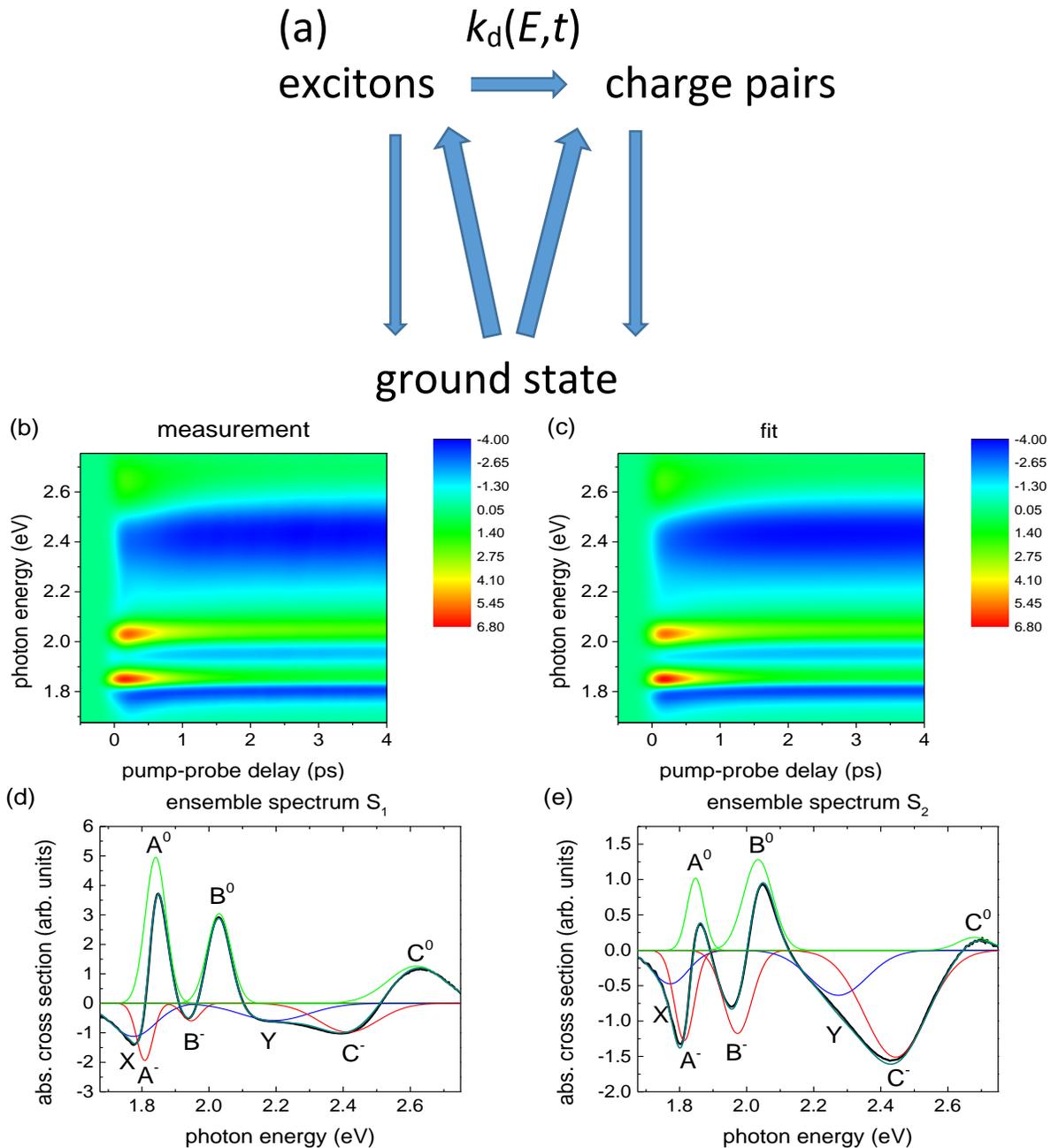

**Figure 4.** (a) Scheme of the photoexcitation dynamics of excitons and charges. (b) contour plot of the fitted $\Delta T/T$ (in %) (c) contour plot of the measured $\Delta T/T$ (in %) (d) spectrum $S_1$ described in the text (black) and fit to eight Gaussian peaks (photoinduced absorption from charges: red, photoinduced absorption from excitons: blue, photobleaching: green) and their sum (dark cyan) (e) spectrum $S_2$ described in the text with the same color coding as (d).

AUTHOR INFORMATION

**Corresponding Authors**

* tetiana.borzda@ijs.si, christoph.gadermaier@ijs.si

REFERENCES

[1] Coleman, J. N. *et al.* Two-Dimensional Nanosheets Produced by Liquid Exfoliation of Layered Materials. *Science* **2011**, 331, 568–571.

[2] Nicolosi, V.; Chhowalla, M.; Kanatzidis, M. G.; Strano, M. S.; Coleman, J. N. Liquid Exfoliation of Layered Materials. *Science* **2013**, 340, 1420.

[3] Wang, Q. H.; Kalantar-Zadeh, K.; Kis, A.; Coleman, J. N.; Strano, M. S. Electronics and optoelectronics of two-dimensional transition metal dichalcogenides. *Nature Nanotech.* **2012**, 7, 699-712.

[4] Butler, S. Z. *et al.* ACS Nano **2013**, 7, 2898-2926

[5] Sipos, B.; Kusmartseva, A. F.; Akrap, A.; Berger, H.; Forro, L.; Tutis, E. From Mott state to superconductivity in $1T$-$TaS_2$. *Nature Mater.* **2008**, 7, 960-965.

[6] Thomson, R.; Burk, B.; Zettl., A.; Clarke, J. Scanning tunneling microscopy of the charge-density-wave structure in $1T$-$TaS_2$. *Phys. Rev. B* **1994**, 49, 16899–16916.

[7] Stojchevska, L.; Vaskivskyi, I.; Mertelj, T.; Kusar, P.; Svetin, D.; Brazovskii, S.; Mihailovic, D. Ultrafast Switching to a Stable Hidden Quantum State in an Electronic Crystal. *Science* **2014**, 344, 177-180.

[8] Radisavljevic, B.; Radenovic, A.; Brivio, J.; Giacometti, V.; Kis, A. Single-layer $MoS_2$ transistors. *Nature Nanotech.* **2011**, 6, 147–150.

[9] Radisavljevic, B.; Whitwick, M. B.; Kis, A. Integrated circuits and logic operations based on single-layer $MoS_2$. *ACS Nano* **2011**, 5, 9934–9938.

[10] Frindt, R. F.; Yoffe A. D. Physical Properties of Layer Structures: Optical Properties and Photoconductivity of Thin Crystals of Molybdenum Disulphide. *Proc. Roy. Soc. A* **1962**, 273, 69-83.

[11] Mak, K. F.; Lee, C.; Hone, J.; Shan, J.; Heinz, T. F. Atomically thin $MoS_2$: a new direct-gap semiconductor. *Phys. Rev. Lett.* **2010**, 105, 136805.

[12] Splendiani, A.; Sun, L.; Zhang, Y.; Li, T.; Kim, J.; Chim, C.-Y.; Galli, G.; Wang, F. Emerging photoluminescence in monolayer $MoS_2$. *Nano Lett.* **2010**, 10, 1271–1275.

[13] Evans, B. L.; Young, P. A. Exciton spectra in thin crystals: the diamagnetic effect. *Proc. Phys. Soc.* **1967**, 91, 475-482.

[14] Ramasubramaniam, A. Large excitonic effects in monolayers of molybdenum and tungsten dichalcogenides. *Phys. Rev. B* **2012**, 86, 115409.

[15] Cheiwchanchamnangij, T.; Lambrecht, W. R. L. Quasiparticle band structure calculation of monolayer, bilayer, and bulk $MoS_2$. *Phys. Rev. B* **2012**, 85, 205302-1-4.

[16] Ye, Z.; Cao, T.; O'Brien, K.; Zhu, H.; Yin, X.; Wang, Y.; Louie, S. G.; Zhang, X. Probing excitonic dark states in single-layer tungsten disulphide. *Nature* **2014**, 513, 214-218.

[17] Fontana, M.; Deppe, T.; Boyd, A. K.; Rinzan, M.; Liu, A. Y.; Paranjape, M.; Barbara, P. Electron Hole Transport and Photovoltaic Effect in Gated MoS2 Schottky Junctions. *Sci.Rep.* **2013**, 3, 1634.

[18] Lopez-Sanchez, O.; Lembke, D.; Kayci, M.; Radenovic, A.; Kis, A. Ultrasensitive Photodetectors Based on Monolayer $MoS_2$. *Nat. Nanotechnol.* **2013**, 8, 497–501.

[19] Mak, K. F.; He, K.; Lee, C.; Lee, G. H.; Hone, J.; Heinz, T. F.; Shan, J. Tightly Bound Trions in Monolayer MoS2. *Nat. Mater.* **2012**, 12, 207–211.

[20] Mouri, S.; Miyauchi, Y.; Matsuda, K. Tunable Photoluminescence of Monolayer MoS2 via Chemical Doping *Nano Lett.* **2013**, 13, 5944-5948.

[21] Scheuschner, N.; Ochedowski, O.; Kaulitz, A. M.; Gillen, R.; Schleberger, M., Maultzsch, J. Photoluminescence of freestanding single- and few-layer $MoS_2$. *Phys. Rev. B* **2014**, 89, 125406.

[22] Dean, N.; Petersen, J. C.; Fausti, D.; Tobey, R. I.; Kaiser, S.; Gasparov, L. V.; Berger, H.; Cavalleri, A.; Polaronic Conductivity in the Photoinduced Phase of $1T$-$TaS_2$. *Phys. Rev. Lett.* **2011** 106, 016401.

[23] Osterbacka, R.; An, C.; Jiang, X.; Vardeny, Z. Two-Dimensional Electronic Excitations in Self-Assembled Conjugated Polymer Nanocrystals. *Science* **2000**, 287, 839−842.

[24] Klimov, V. I.; Ivanov, S. A.; Nanda, J.; Achermann, M.; Bezel, I.; McGuire, J. A.; Piryatinski, A. Single-exciton optical gain in semiconductor nanocrystals *Nature* **2007**, 447, 441−446.

[25] Cabanillas-Gonzalez, J.; Virgili, T.; Gambetta, A.; Lanzani, G.; Anthopoulos, T. D.; de Leeuw, D. M. *Phys. Rev. Lett.* **2006**, 96, 106601

[26] Gadermaier, C.; Menna, E.; Meneghetti, M.; Kennedy, W. J.; Vardeny, Z. V.; Lanzani, G. Long-Lived Charged States in Single-Walled Carbon Nanotubes. *Nano Lett.* **2006**, 6, 301−305.

[27] Gatensby, R.; McEvoy, N.; Lee, K.; Hallam, T.; Berner, N. C.; Rezvani, E.; Winter, S.; O'Brian, M.; Duesberg, G. S. Controlled synthesis of transition metal dichalcogenide thin films for electronic applications. *Appl. Surf. Sci.* **2014**, 297, 139-146

[28] Lee, C.; Yan, H.; Brus, L. E.; Heinz, T. F.; Hone, J.; Ryu, S.; Anomalous Lattice Vibrations of Single- and Few-layer $MoS_2$. *ACS Nano* **2010**, 4, 2695-2700.

[29] Li, H.; Zhang, Q.; Yap, C. C. R.; Tay, B. K.; Edwin, T. H. T.; Olivier, A.; Baillargeat, D. From Bulk to Monolayer $MoS_2$: Evolution of Raman Scattering. *Adv. Funct. Mater.* **2012**, 22, 1385-1390.

[30] Shi, H.; Yan, R.; Bertolazzi, S.; Brivio J.; Bao, G.; Kis, A.; Jena, D., Xing, H. G.; Huang, L. Exciton Dynamics in Suspended Monolayer and Few-Layer $MoS_2$ 2D Crystals. *ACS Nano* **2013**, 7, 1072-1080.

[31] Mai, C.; Barrette, A.; Yu, Y.; Semenov, Yu. G.; Kim, K. W.; Cao, L.; Gundogdu, K. Many-Body Effects in Valleytronics: Direct Measurement of Valley Lifetimes in Single-Layer $MoS_2$. *Nano Lett.* **2014**, 14, 202-206.

[32] Sercombe, D., Schwarz, S., Del Pozo-Zamudio, O.; Liu, F.; Robinson, B. J.; Chekhovich, E. A.; Tartakovskii, I. I.; Kolosov, O.; Tartakovskii, A. I. Optical investigation of the natural electron doping in thin $MoS_2$ films deposited on dielectric substrates. *Sci. Rep.* **2013**, 3, 3489.

[33] Plechinger, G.; Schrettenbrunner, F.-X.; Eroms, J.; Weiss, D.; Schuller, C.; Korn, T. Low-temperature photoluminescence of oxide-covered single-layer $MoS_2$. *Phys. Stat. Sol. RRL* **2012**, 6, 126-128.

[34] Gutierrez, H. R. et al. Extraordinary Room-Temperature Photoluminescence in Triangular $WS_2$ Monolayers. *Nano Lett.* **2013**, 13, 3447-3454.

[35] Yadgarov, L.; Choi, C. L.; Sedova, A.; Cohen, A.; Rosentsveig, R.; Bar-Elli, O.; Oron, D., Dai, H.; Tenne, R. Dependence of the Absorption and Optical Surface Plasmon Scattering of $MoS_2$ Nanoparticles on Aspect Ratio, Size, and Media. *ACS Nano* **2014**, 8, 3575-3583.

[36] Soavi, G.; Scotognella, F.; Brida, D.; Hefner, T.; Späth, F.; Antognazza, M. R.; Hertel, T.; Lanzani, G.; Cerullo, G. Ultrafast Charge Photogeneration in Semiconducting Carbon Nanotubes. *J. Phys. Chem. C* **2013**, 117, 10849-10855.

[37] Sim, S.; Park, J.; Song, J.-G.; In, C.; Lee, Y.-S.; Kim, H.; Choi, H. Exciton dynamics in atomically thin $MoS_2$: Interexcitonic interaction and broadening kinetics. *Phys. Rev. B* **2013**, 88, 075434-1-5.

[38] Tränkle, G.; Lach, E.; Forchel, A.; Scholz, F.; Ell, C.; Haug, H.; Weimann, G.; Griffiths, G.; Kroemer, H.; Subbanna, S. General Relation between Band-Gap Renormalization and Carrier Density in Two-Dimensional Electron-Hole Plasmas. *Phys. Rev. B* **1987**, 36, 6712-6714.

[39] Wang, Q.; Ge, S.; Li, X.; Qiu, J.; Ji, Y.; Feng, J.; Sun, D. Valley Carrier Dynamics in Monolayer Molybdenum Disulfide from Helicity-Resolved Ultrafast Pump-Probe Spectroscopy. *ACS Nano* **2013**, 7, 11087-11093.

[40] Ugeda, M. M. et al. Giant bandgap renormalization and excitonic effects in a monolayer transition metal dichalcogenide semiconductor. *Nature Mater.* advance online publication.

[41] Ouyang, Q. Y.; Yu, H. L.; Zhang, K.; Chen, Y. J. Saturable absorption and the changeover from saturable absorption to reverse saturable absorption of $MoS_2$ nanoflake array films. *J. Mater Chem C* **2014**, 2, 6319-25.

[42] Kozawa, D. et al. Photocarrier relaxation pathway in two-dimensional semiconducting transition metal dichalcogenides. *Nature Commun.* **2014**, 5, 4543-1-5.

[43] Gadermaier, C.; Cerullo, G.; Sansone, G.; Leising, G.; Scherf, U.; Lanzani, G. Time-Resolved Charge Carrier Generation from Higher Lying Excited States in Conjugated Polymers. *Phys. Rev. Lett.* **2002**, 89, 117402.

[44] Noolandi, J.; Hong, K. M. Theory of photogeneration and fluorescence quenching. *J. Chem. Phys.* 1979, 70 3230-3236.

[45] Arkhipov, V. I., Emilianova, E. V., Bassler, H. Hot exciton dissociation in a conjugated polymer. *Phys. Rev. Lett.* **1999**, 82, 1321-1324.

[46] Leitenstorfer, A.; Furst, C.; Laubereau, A.; Kaiser, W.; Trankle, G.; Weimann, G. Femtosecond carrier dynamics in GaAs far from equilibrium. *Phys. Rev. Lett.* **1996**, 76, 1545-1548.

[47] Sun, D.; Wu, Z.-K.; Divin, C.; Li, X.; Berger, C.; de Heer, W. A.; First, P. N., Norris, T. B.; Ultrafast Relaxation of Excited Dirac Fermions in Epitaxial Graphene Using Optical Differential Transmission Spectroscopy. *Phys. Rev. Lett.* **2008**, 101, 157402-1-4.

[48] Gadermaier, C. *et al.* Strain-Induced Enhancement of the Electron Energy Relaxation in Strongly Correlated Superconductors. *Phys. Rev. X* **2014**, 4, 011056-1-6.

[49] Dhar, L.; Roger, J. A.; Nelson, K. A. Time-Resolved Vibrational Spectroscopy in the Impulsive Limit. *Chem. Rev.* **1994**, 94, 157-193.

[50] Mai, C.; Barrette, A.; Yu, Y. F.; Semenov, Y. G.; Kim K. W.; Cao, L. Y.; Gudnogdu, K. Many-Body Effects in Valleytronics: Direct Measurement of Valley Lifetimes in Single-Layer $MoS_2$. *Nano Lett.* **2014**, 14, 202-206.

[51] Lagarde, D.; Bouet, L.; Marie, X.; Zhu, C. R.; Liu, B. L.; Amand, T.; Tan, P. H.; Urabaszek, B. Carrier and Polarization Dynamics in Monolayer $MoS_2$. *Phys. Rev. Lett.* **2014**, 114, 047401-1-4.

[52] Assael, M. J.; Botsios, S.; Gialou, J.; Metaxa, I. N. Thermal Conductivity of Polymethyl Methacrylate (PMMA) and Borosilicate Crown Glass BK7. *Int. J. Thermophys.* **2005**, 26, 1595-1605.

[53] Sun, Q. C.; Yadgarov, L.; Rosentsveig, R.; Seifert, G.; Tenne, R.; Musfeldt, J. L.; Observation of a Burstein-Moss Shift in Rhenium-Doped $MoS_2$ Nanoparticles. *ACS Nano* **2013**, 7, 3506-11.

[54] Pospischil, A.; Furchi, M. M.; Mueller, T. Solar-energy conversion and light emission in an atomic monolayer p-n diode. *Nature Nanotech* **2014**, 9, 257-261.

[55] Bauger, B. W. H.; Churchill, H. O. H.; Yang, Y. F., Jarillo-Herrero, P. Optoelectronic devices based on electrically tunable p-n diodes in a monolayer dichalcogenide. *Nature Nanotech.* **2014**, 9, 262-267.

[56] Britnell, L. *et al*. Strong Light-Matter Interactions in Heterostructures of Atomically Thin Films. *Science* **2013**, 340, 1311-1314.

[57] Lee, C.-H. et al. Atomically thin p-n junctions with van der Waals heterointerfaces. *Nature Nanotech.* **2014**, 9, 676-681.

[58] Hong, X.; Kim, J.; Shi, S.-F.; Zhang, Y.; Jin, C.; Sun, Y.; Tongay, S.; Wu, J.; Zhang, Y.; Wang, F. Ultrafast charge transfer in atomically thin $MoS_2$/$WS_2$ heterostructures. *Nature Nanotech.* **2014**, 9, 682-686.